\documentclass[10pt,floatfix,nofootinbib,aps,prc,twocolumn]{revtex4}

\usepackage[dvips]{graphicx}
\usepackage{amsmath,amssymb,amsopn,bm}
\usepackage{color}
\usepackage{epstopdf}

\newcommand{\dd}{\text{d}}

\newcommand{\beq}{\begin{equation}}
\newcommand{\eeq}{\end{equation}}
\newcommand{\bce}{\begin{center}}
\newcommand{\ece}{\end{center}}
\newcommand{\eg}{\textit{e.g.}}

\newcommand{\lsim}{\lesssim}
\newcommand{\gsim}{\gtrsim}

\newcommand{\MeV}{\ensuremath{\mathrm{MeV}}}
\newcommand{\GeV}{\ensuremath{\mathrm{GeV}}}
\newcommand{\fm}{\ensuremath{\mathrm{fm}}}

\begin{document}

\title{Thermal Photons and Collective Flow at the Relativistic Heavy-Ion
Collider} 

\author{Hendrik van Hees}
\affiliation{Institut f{\"u}r Theoretische Physik, Goethe-Universit{\"a}t
Frankfurt, Germany, \linebreak Frankfurt Institute for Advanced Studies (FIAS),
Ruth-Moufang-Str.~1, D-60438 Frankfurt, Germany}

\author{Charles Gale}
\affiliation{Department of Physics, McGill University, 3600 University
Street, Montreal, Canada H3A 2T8}

\author{Ralf Rapp} 
\affiliation{Cyclotron Institute and Department of Physics\,\&\,Astronomy, 
Texas A{\&}M University, College Station, Texas 77843-3366, USA }

\date{\today}

\begin{abstract} 
  We update our calculations of thermal-photon production in nuclear
  collisions at the Relativistic Heavy-Ion Collider
  (RHIC). Specifically, we address the recent experimental observation
  of an elliptic flow of direct photons comparable in magnitude to that
  of pions, which is at variance with expectations based on quark-gluon
  plasma (QGP) dominated photon radiation. Our thermal emission rate is
  based on previous work, i.e., resummed leading-order QGP emission and
  in-medium hadronic rates in the confined phase.  These rates are
  nearly degenerate at temperatures close to the expected QCD-phase
  change. The rates are convoluted over an improved elliptic-fireball
  expansion with transverse- and elliptic-flow fields quantitatively
  constrained by empirical light- and strange-hadron spectra. The
  resulting direct-photon spectra in central Au-Au collisions are
  characterized by hadron-dominated emission up to transverse momenta of
  $\sim 2$-$3 \; \GeV$. The associated large elliptic flow in the
  hadronic phase mitigates the discrepancy with the measured
  photon-$v_2$ compared to scenarios with QGP-dominated emission.
\end{abstract}

\maketitle

\section{Introduction}  
\label{sec_intro}

Dileptons and photons are the only particles which escape the interior
of the fireball created in ultrarelativistic heavy-ion collisions
(URHICs) unaffected. Their production, on the other hand, is rooted in
the strongly interacting medium and thus illuminates the latter's
properties, see Refs.~\cite{Rapp:2009yu,Gale:2009gc,Alam:1999sc} for
reviews. Recent highlights of electromagnetic observables in URHICs
include ``measurements'' of the in-medium vector-spectral function via
dilepton \emph{invariant}-mass spectra at the
SPS~\cite{Arnaldi:2006jq,Adamova:2006nu,Arnaldi:2008fw} and of the
medium temperature via direct photons at RHIC~\cite{Adare:2008fq}.

It is well known that the thermal component in the observed
electromagnetic spectra results from a convolution of the temperature-
and density-dependent emission rate over the entire space-time history
of the expanding fireball in URHICs. From the interplay of decreasing
temperature and increasing three-volume in the course of the fireball
expansion, it can be fairly well established that photon and dilepton
emission in the low-energy regime, $q_0\lsim 1\; \GeV$\footnote{The
  energy variable, $q_0$, encompasses both mass ($M$) and
  transverse-momentum ($q_t$) dependencies, \eg, $q_0=(M^2+q_t^2)^{1/2}$
  for $q_z=0$.}, are dominated by the hadronic phase. At energies beyond
$\sim 1\;\GeV$, the situation is less clear, since the competition
between hadronic and QGP sources will be sensitive to additional
ingredients~\cite{vanHees:2007th}, \eg, the relative strength of
hadronic and QGP emission rates, the phase-transition temperature (which
formally demarcates the space-time dependence of the two sources) and
the transverse flow (inducing a blue shift to higher $q_t$ which is more
pronounced in the later hadronic phase)\footnote{Note that the
  blue-shift distortion does not apply to dilepton \emph{invariant}-mass
  spectra if a finite detector acceptance can be fully corrected for.}.

In this context, recent measurements of the elliptic flow of direct
photons (i.e., after the subtraction of long-lived hadron decays, mostly
$\pi^0,\eta\to\gamma\gamma$) in semicentral Au-Au
collisions~\cite{Adare:2011zr} have revealed remarkable results. It has
been found that the pertinent flow coefficient, $v_2^{\gamma}(q_t)$, is
as large as that of charged pions up to momenta of $q_t\simeq 3 \;\GeV$
(albeit the photon data carry somewhat larger error bars). This result
is difficult to reconcile with a dominant QGP emission
source. Primordial photons from binary $N$-$N$ collisions, whose
emission is expected to be isotropic, will further reduce the total
direct-photon $v_2$. Current model
calculations~\cite{Chatterjee:2006,Liu:2009,Holopainen:2011pd} using a
hydrodynamically expanding medium with QGP and hadronic radiation, as
well as primordial photons, underpredict the experimentally measured
$v_2$ by a factor of $\sim 5$ ($\sim 3$ when accounting for the maximal
systematic error in the measurement). The question thus arises what
could be missing in these calculations.

In the present paper we re-examine several aspects related to thermal
photon emission from the hadronic medium. Since the $v_2$ in the
hadronic phase of URHICs is large, an augmented hadronic component is a
natural candidate to improve the description of thermal-photon emission
at RHIC and thus reduce the discrepancy with the $v_2$ data. First, we
note that typical hydrodynamical evolutions with first-order phase
transitions tend to underestimate the radial (and possibly elliptic)
flow built up in the fireball at the end of the QGP phase. This was
borne out of a recent phenomenological analysis of light- and
strange-hadron spectra using blast-wave parameterizations of the
respective sources at thermal and chemical and freezeout (with
$T_{\mathrm{ch}}\simeq 100 \; \MeV$ and $T_{\mathrm{ fo}}\simeq180 \;
\MeV$, respectively)~\cite{He:2010vw}. In particular, the observed
universality in kinetic-energy scaling of the $v_2$ of light and strange
hadrons, together with an earlier decoupling of multi-strange hadrons
(as inferred from their $p_t$ spectra), suggests that most of the
hadronic $v_2$ is indeed of partonic
origin~\cite{Abelev:2007rw}. Second, for the thermal emission rates from
hadronic matter we use our previous results of
Ref.~\cite{Turbide:2003si}, which, in particular, include contributions
from baryons which are known to be important from dilepton
calculations~\cite{vanHees:2007th}, even at RHIC~\cite{Rapp:2000pe}.
Finally, we take into account chemical off-equilibrium effects in the
hadronic phase, i.e., effective chemical potentials for pions, kaons,
etc., which can further augment the hadronic component in thermal-photon
spectra (\eg, typical processes like $\pi\rho\to\pi\gamma$ are enhanced
by a pion fugacity to the third power).

Our paper is organized as follows. In Sec.~\ref{sec_sources} we briefly
recall our input for the thermal-photon rates as taken from
Ref.~\cite{Turbide:2003si}, as well as for the primordial contribution,
which we check against $pp$ data. In Sec.~\ref{sec_med} we update our
description of the thermal fireball evolution by constructing (a time
evolution of) flow fields which is consistent with the empirical
extraction at chemical and thermal freezeout. In Sec.~\ref{sec_pt-spec}
we evolve the rates over the fireball evolution and discuss the
resulting photon-$q_t$ spectra, in particular the composition of the
thermal yields. In Sec.~\ref{sec_v2} we present our results for the
direct-photon $v_2$ in comparison to the recent PHENIX
data~\cite{Adare:2011zr} and in light of other model results. Finally,
Sec.~\ref{sec_concl} contains our conclusions.

\section{Photon Sources}
\label{sec_sources}

The photon spectrum resulting from a heavy-ion (or $pp$) reaction is
usually referred to as inclusive photons. The subtraction of long-lived
final-state decays leads to the notion of direct photons, which are the
ones of interest in the present context. For direct photons, we further
distinguish the radiation of thermal photons (Sec.~\ref{ssec_thermal}),
characterized by an equilibrium-emission rate to be integrated over the
space-time evolution of the medium, and a non-thermal component
emanating from primordial interactions prior to thermalization
(Sec.~\ref{ssec_nontherm}).

\subsection{Thermal Emission}
\label{ssec_thermal}

In the present work we adopt the thermal emission rates of photons as
developed and compiled in Ref.~\cite{Turbide:2003si}.

For QGP radiation we use the numerical parameterization of the complete
leading-order in $\alpha_s$ rate as given in Ref.~\cite{Arnold:2001ms}.
The main input for the QGP rate is the strong coupling ``constant'' for
which we take an expression with temperature-dependent one-loop running
at the scale $\sim2\pi T$, $\alpha_s(T)=6\pi/27\ln(T[\GeV]/0.022)$.
This amounts to values of around 0.3 in the relevant temperature regime,
$T=1$-$2 T_c$, which turns out to be consistent with recent estimates
from the (perturbative) Coulomb term in in-medium heavy-quark free
energies~\cite{Riek:2010fk}.

The basis of the thermal emission rate in hadronic matter forms the
electromagnetic correlation function computed in
Refs.~\cite{Rapp:1999us,Turbide:2003si} using hadronic many-body theory
with effective Lagrangians. It has been successfully
used~\cite{Rapp:2009yu} in the interpretation of dilepton data at the
SPS~\cite{Arnaldi:2006jq,Adamova:2006nu} and has been carried to the
photon point in Ref.~\cite{Turbide:2003si}. It includes a rather
extensive set of meson and baryon resonances in the interaction of the
isovector current with a thermal heat bath of hadrons. It has been
augmented by additional meson-exchange reactions in a meson gas which
become important at photon momenta $q\gsim 1 \; \GeV$, \eg, $\pi$,
$\omega$, and $a_1$ exchange in $\pi+\rho\to\pi+\gamma$ as well as
strangeness-bearing reactions (\eg, $\pi+K^*\to
K+\gamma$)~\cite{Turbide:2003si}. An important element in constraining
the hadronic vertices to empirical information, such as hadronic and
radiative decay branchings, is the (gauge-invariant) introduction of
vertex form factors. The latter lead to a substantial reduction of the
hadronic emission rate with increasing photon momentum, which is
essential for quantitative descriptions of hadronic emission rates at
the momenta of experimental interest ($q_t\lsim 3.5 \; \GeV$ for thermal
radiation). Without vertex form factors, hadronic photon rates should
not be considered reliable for momenta $q\geq1 \; \GeV$.  We do not
include here additional $\pi\pi$ Bremsstrahlung contributions
($\pi\pi\to\pi\pi\gamma$) as evaluated in
Refs.~\cite{Srivastava:2004xp,Liu:2007zzw}. This source is important at
low momenta, $q_t \lsim 0.5 \; \GeV$, put plays no role in the region of
interest of the present investigation ($q_t\ge1 \; \GeV$).

The resulting total hadronic and QGP emission rates were found to be
remarkably close to each other for temperatures around the putative
transition temperature, $T_c\simeq180 \; \MeV$. This is appealing from a
conceptual point of view in terms of a possibly continuous matching of
the bottom-up and top-down extrapolated hadronic and QGP rates,
respectively. It is also welcome from a practical point of view, since
it much reduces the uncertainties associated with identifying the medium
in the fireball as hadronic or partonic, which, in the case of a
cross-over, may not even be well defined. If QGP and hadronic rates are
significantly different across $T_c$, appreciable uncertainties in the
calculated photon spectra have been found as a result of this
ambiguity~\cite{Holopainen:2011pd}.

\subsection{Non-Thermal Sources}
\label{ssec_nontherm}

\begin{figure}[!t]
\begin{center}
\begin{minipage}{0.9\linewidth}
\includegraphics[width=\textwidth]{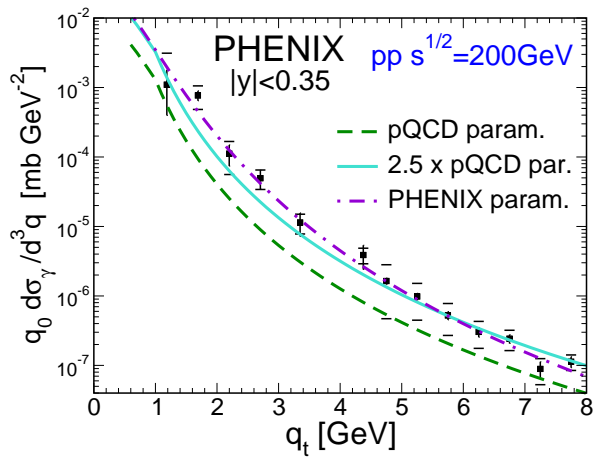}
\end{minipage} 
\end{center}
\caption{(Color online) Empirical fits to the direct-photon spectrum
  measured by PHENIX~\cite{Adler:2006yt} in $p$-$p$ collisions at
  $\sqrt{s} =200\; \GeV$; dash-dotted line: power-law
  fit~\cite{Adler:2006yt}, solid and dashed line: $x_t$-scaling
  ansatz~\cite{Srivastava:2001bw} with and without $K$ factor,
  respectively.}
\label{fig_pQCD} 
\end{figure}

After subtraction of final-state decays, the main source other than
``thermal'' radiation from the interacting medium is associated with
``primordial'' photons produced upon first impact of the nuclei via
binary $NN$ collisions. Primordial photon spectra are usually estimated
from the direct contribution in $pp$ collisions, where no significant
reinteractions are expected. This is supported by the generally good
agreement of the measured photon spectra with next-to-leading order
(NLO) perturbative QCD (pQCD) calculations~\cite{Gordon:1993qc} for
primordial production. In the following we will adopt a simple power-law
fit performed by the PHENIX
collaboration~\cite{Adare:2008fq,David:2011priv} to their $pp$
data~\cite{Adler:2006yt,Adare:2008fq}, cf.~Fig.~\ref{fig_pQCD}.
Figure~4 in Ref.~\cite{Adare:2008fq} reveals that the NLO pQCD
calculations are slightly below the PHENIX parameterization in the
relevant $q_t$ range of $\sim 1$-$ 7 \; \GeV$. Therefore, we will
alternatively estimate the primordial contribution via a $x_t$-scaling
motivated parameterization~\cite{Srivastava:2001bw}, upscaled by a
$K$-factor of 2.5 to best match the PHENIX $pp$ data for $q_t=5$-$8 \;
\GeV$.  This fit produces a slightly smaller yield of photons with
momenta $q_t=1$-$5\;\GeV$ compared to the PHENIX fit and is thus very
similar to the NLO pQCD calculations.

In principle, photons can also be radiated off fast-moving partons
(jets) interacting with the medium. The combined elliptic flow of this
source is expected to be close to zero~\cite{Turbide:2005bz}. We
therefore subsume this source in our primordial contribution (whose
fragmentation part, \eg, is expected to be reduced in the medium), which
is a posteriori justified by an adequate description of the spectral
yields in Au-Au collisions once thermal radiation is implemented.

\section{Fireball and Transverse-Flow Field}
\label{sec_med}

\begin{figure*}[!t]
\begin{center}
\begin{minipage}{0.48\linewidth}
\includegraphics[width=\textwidth]{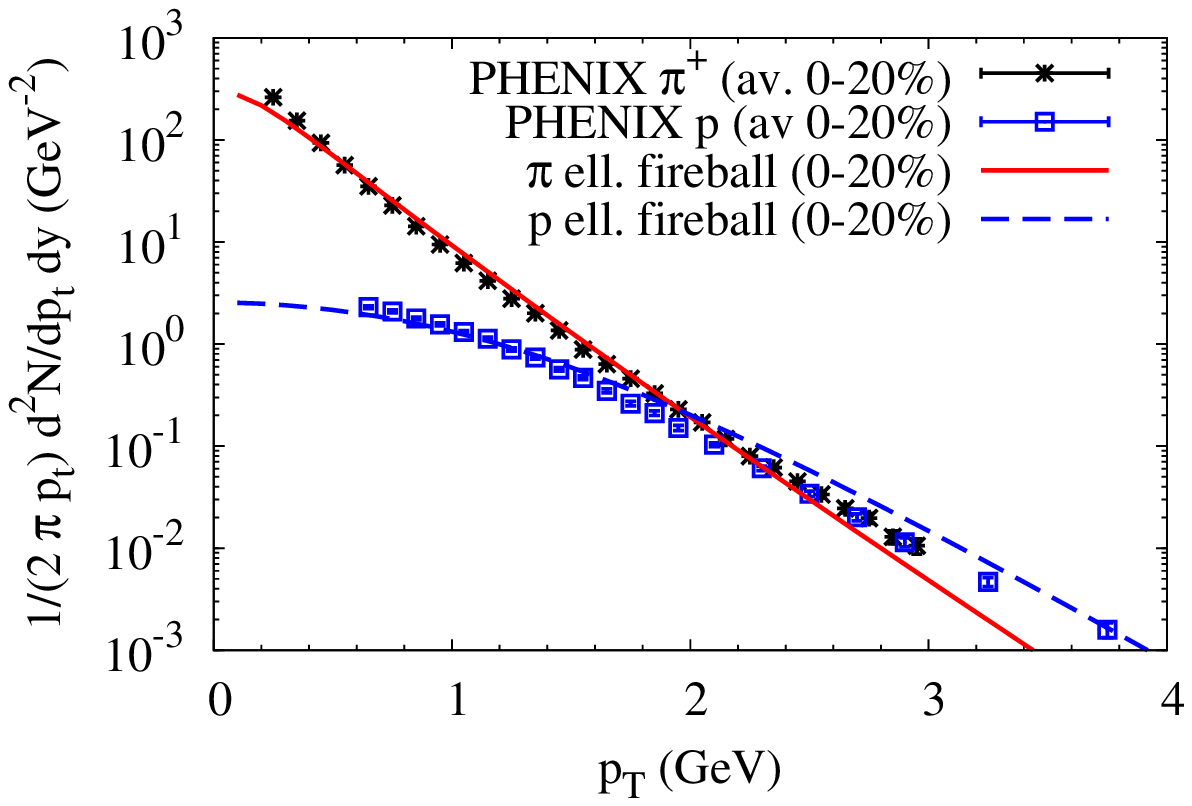}
\end{minipage} \hfill
\begin{minipage}{0.48\linewidth}
\includegraphics[width=\textwidth]{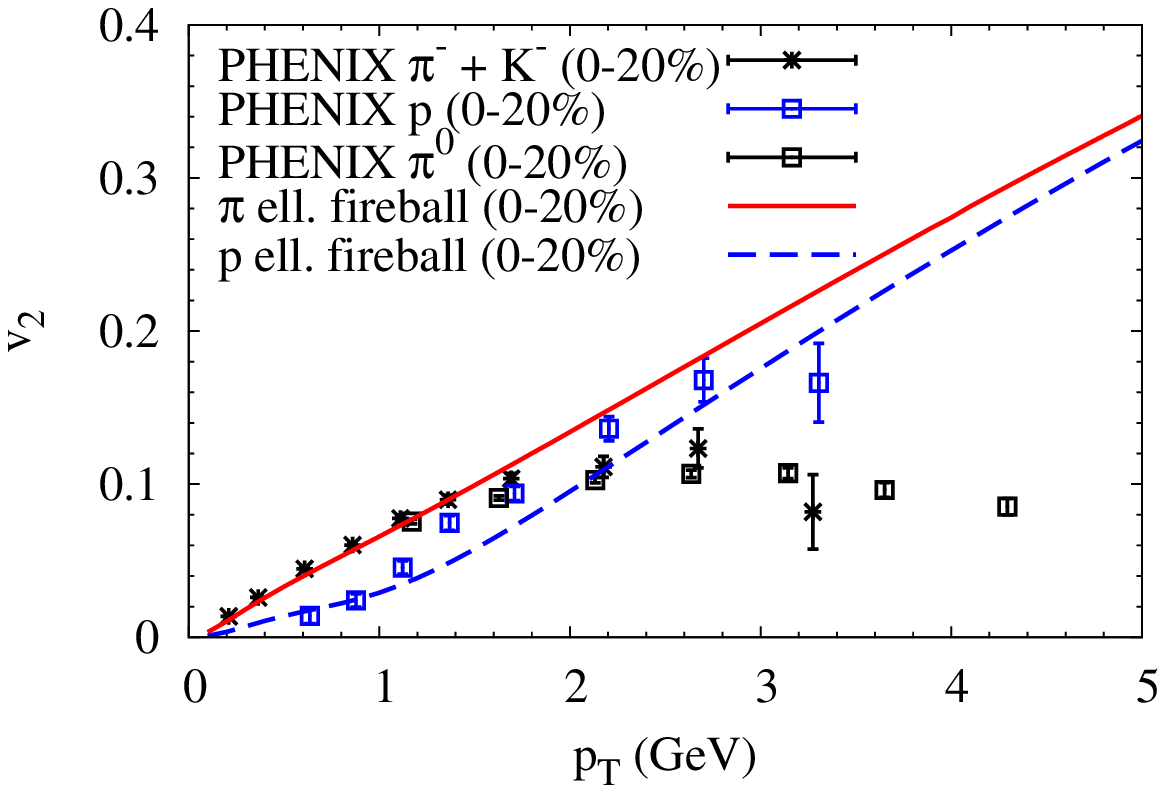} 
\end{minipage}
\begin{minipage}{0.48\linewidth}
\includegraphics[width=\textwidth]{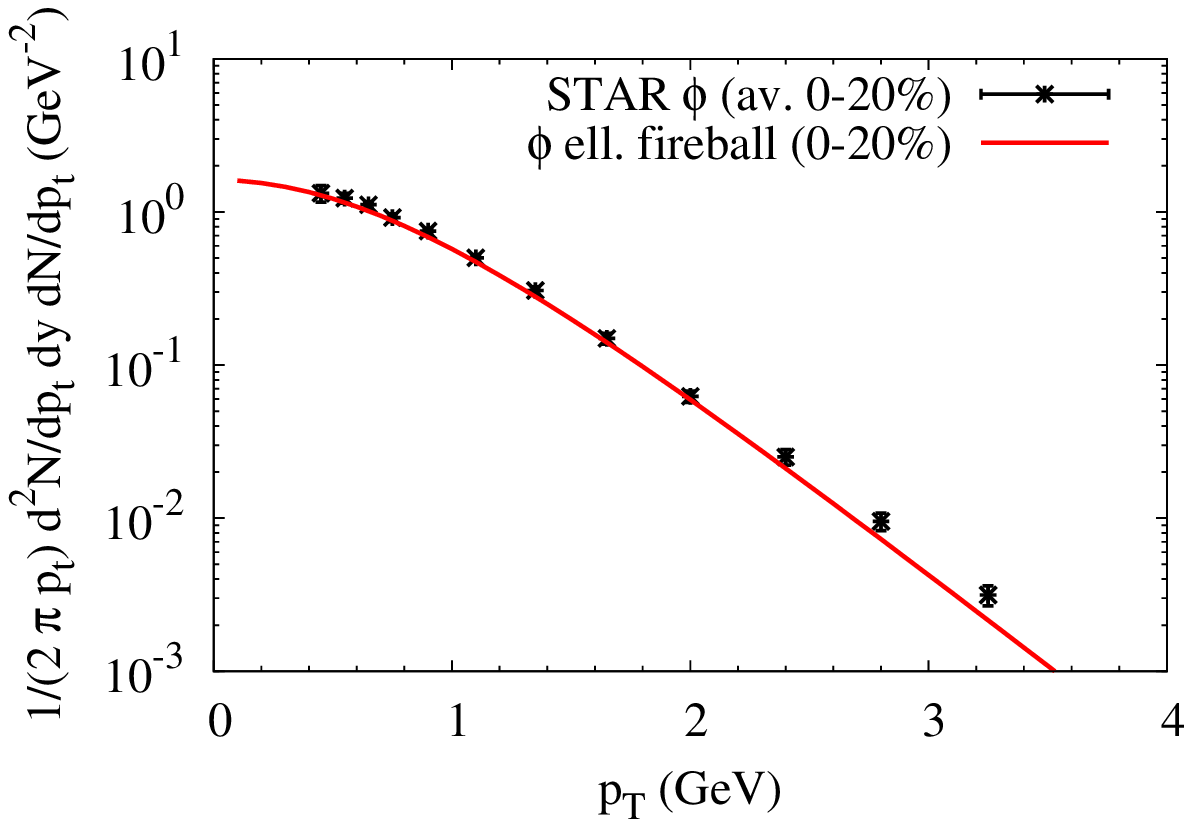}
\end{minipage} \hfill
\begin{minipage}{0.48\linewidth}
\includegraphics[width=\textwidth]{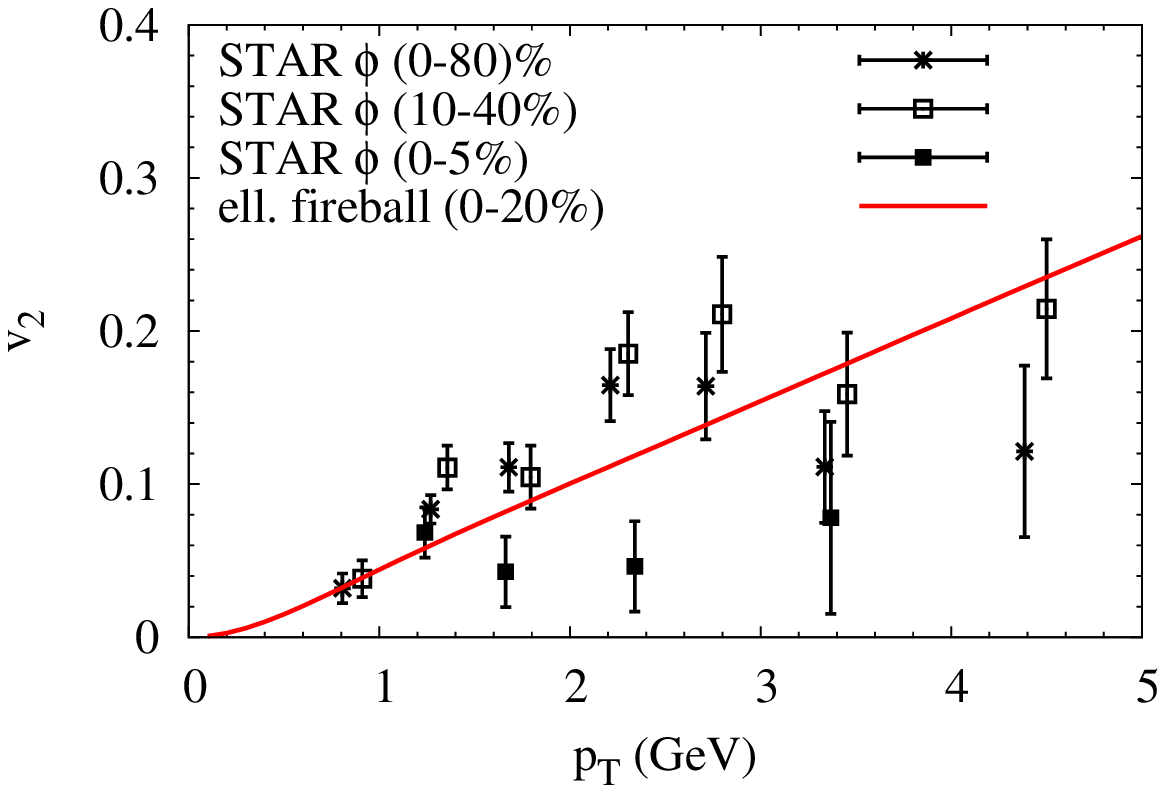}
\end{minipage} 
\end{center}
\caption{(Color online) Snapshots of $p_t$ spectra and $v_2$ for pions
  and protons (upper panels), as well as $\phi$ mesons (lower panels),
  following from our fireball evolution in 0-20\% Au-Au($\sqrt{s} =200\;
  A \GeV$) collisions at thermal and chemical freezeout,
  respectively. The $\pi$ and $p$ curves are for direct emission only
  (no resonance feeddown) with absolute normalization while the $\phi$
  yield is (re-) normalized to the data. Data are from
  Refs.~\cite{Adler:2003kt,Adler:2003cb,Abelev:2007rw}.}
\label{fig_hadrons}
\end{figure*}

The continuous emission of photons throughout the evolution of a
heavy-ion reaction causes their elliptic-flow signal to be more
sensitive to its time evolution than that of hadronic final states.
Therefore, a calculation of the photon-$v_2$ requires special care in
constructing a realistic time evolution of both radial and elliptic flow
(the former affects the spectral weight of photon emission at a given
time snapshot). In principle, hydrodynamic models are believed to be
able to accomplish such a task; however, current uncertainties including
initial conditions (\eg, fluctuations and initial flow fields),
viscosity corrections and the coupling to hadronic cascades in the
dilute stages render this a challenging task, which has not been
completed (yet). In the present paper we take a more pragmatic (and
simple) approach which nevertheless accurately captures two
experimentally established snapshots of the fireball evolution, namely
the flow fields at chemical freezeout at $T_{\mathrm{ch}} \simeq 170 \;
\MeV$ and kinetic freezeout at $T_{\mathrm{fo}}\simeq 100 \; \MeV$. The
latter is well determined by the transverse-momentum ($p_t$) spectra and
elliptic flow of light hadrons ($\pi$, $K$,
$p$)~\cite{Adare:2006ti,Abelev:2008ed,Abelev:2008ez}, while the former
can be extracted from spectra and flow of multi-strange
hadrons\cite{Abelev:2007rw,Afanasiev:2007tv,Abelev:2008ed}. A detailed
fit of these snapshots using an elliptic blast-wave source has been
performed in Ref.~\cite{He:2010vw}, which was also shown to be
compatible with the empirical constituent-quark number and transverse
kinetic-energy (KE$_T$) scaling of the elliptic flow of light and
strange hadrons. We use these results to improve a previously
constructed expanding elliptic fireball~\cite{vanHees:2005wb} so that
its evolution passes through these benchmarks at the end of the mixed
phase and at thermal freezeout.  Representative examples of the
resulting multi-strange ($\phi$ mesons) and light-hadron ($\pi$, $p$)
spectra and $v_2$ at chemical and thermal freezeout, respectively, are
illustrated in Fig.~\ref{fig_hadrons}.

Concerning initial conditions, we assume an initial longitudinal
fireball size of $z_0=0.6 \;\fm$ corresponding to $c\tau_0\simeq z_0/
\Delta y \simeq0.33 \; \fm$ ($\Delta y\simeq1.8$) as our default value
for both 0-20\% and 20-40\% centrality classes. With total entropies of
$S=7900$ and 3600 (assumed to be conserved), this translates into
(average) initial tempe\-ratures of $T_0\simeq355 \; \MeV$ and $325\;
\MeV$, and charged-hadron multiplicities of $\dd N_{\mathrm{ch}}/\dd
y\simeq610$ and 280, respectively, adjusted to recent STAR
data~\cite{Abelev:2008ez}.

\begin{figure}[!t]
\begin{center}
\begin{minipage}{0.95\linewidth}
\includegraphics[width=\textwidth]{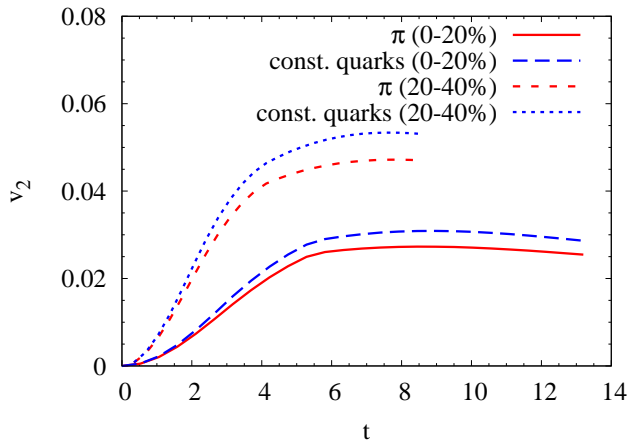}
\end{minipage} \hfill
\end{center}
\caption{(Color online) Time evolution of the inclusive elliptic flow
  for 0-20\% and 20-40\% Au-Au($\sqrt{s}$=200\,AGeV) collisions within
  our fireball model, evaluated with either constituent-quark or pion
  content of the medium.}
\label{fig_v2-tau}
\end{figure}
The time evolution of the inclusive elliptic flow is shown in
Fig.~\ref{fig_v2-tau}. The observed $\mathrm{KE}_T$ scaling of light and
multi-strange hadrons is the main constraint which requires a rather
rapid increase of the bulk $v_2$, and a subsequent leveling off shortly
after chemical freezeout. The final values of $\sim$2.5\% and $\sim$5\%
for the pion-$v_2$ are adjusted to experimental
data~\cite{Abelev:2008ed}.

Another important aspect in our medium evolution is the implementation
of chemical freezeout, i.e., the use of effective meson (and baryon)
chemical potentials to preserve the observed hadron ratios in the
fireball expansion at temperatures between $T_{\mathrm{ch}}$ and
$T_{\mathrm{fo}}$.  We do this as described in Ref.~\cite{Rapp:2002fc},
which was adopted in our previous work~\cite{Turbide:2003si}. Most of
the hydrodynamic evolutions used for photon calculations at RHIC to date
assume chemical equilibrium throughout the hadronic phase. This
assumption likely leads to an appreciable underestimate of the thermal
hadronic component in the observed photon spectra, and thus of its
contribution to the direct-photon elliptic flow. For example, typical
meson annihilation processes such as $\pi+\rho\to\pi+\gamma$ (proceeding
through $t$- and $s$-channel $\pi$, $\omega$ and $a_1$ exchanges), are
augmented by an initial pion fugacity, $z_\pi^3 =\exp(3\mu_\pi/T)$ (in
Boltzmann approximation), where $\mu_\pi\simeq 100 \; \MeV$ in the
vicinity of thermal freezeout, $T_{\mathrm{fo}}\simeq100\; \MeV$. This
implies a significantly larger enhancement in photon production in the
later hadronic stages relative to the conservation of the hadron ratios
for which the chemical potentials are introduced. In other words, the
faster cooling of the fireball in chemical off-equilibrium relative to
the equilibrium evolution is overcompensated in the leading
photon-production channels due to a ``high'' power of pion densities.

\section{Direct-Photon Spectra}
\label{sec_pt-spec}

We start the comparison of our theoretical calculations of direct
photons to data at RHIC with the absolute yields in the
transverse-momentum ($q_t$ spectra).  Let us first illustrate the
quantitative effect of updating the radial expansion starting from our
original predictions in Ref.~\cite{Turbide:2003si}.\footnote{For
  simplicity we will use for this purpose a cylindrically symmetric
  fireball (no $v_2$) and apply an average boost of 70\% of the fireball
  surface flow to the photon spectra in the rest frame,
  $\langle\beta\rangle = 0.7 \beta_{\mathrm{s}}$.}  In the latter, a
transverse acceleration of the fireball surface of $a_T=0.053c^2/\fm$
had been assumed, which, together with a fireball lifetime of
$15\;\fm/c$, leads to a surface velocity of $\beta_s\simeq0.62$ and a
freezeout temperature of $T_{\mathrm{fo}}=108\; \MeV$ for Au-Au
collisions in the 0-20\% centrality bin ($N_{\mathrm{part}}=280$ and
$N_{\mathrm{coll}}=765$).
\begin{figure}[!t]
\begin{center}
\begin{minipage}{0.9\linewidth}
\includegraphics[width=\textwidth]{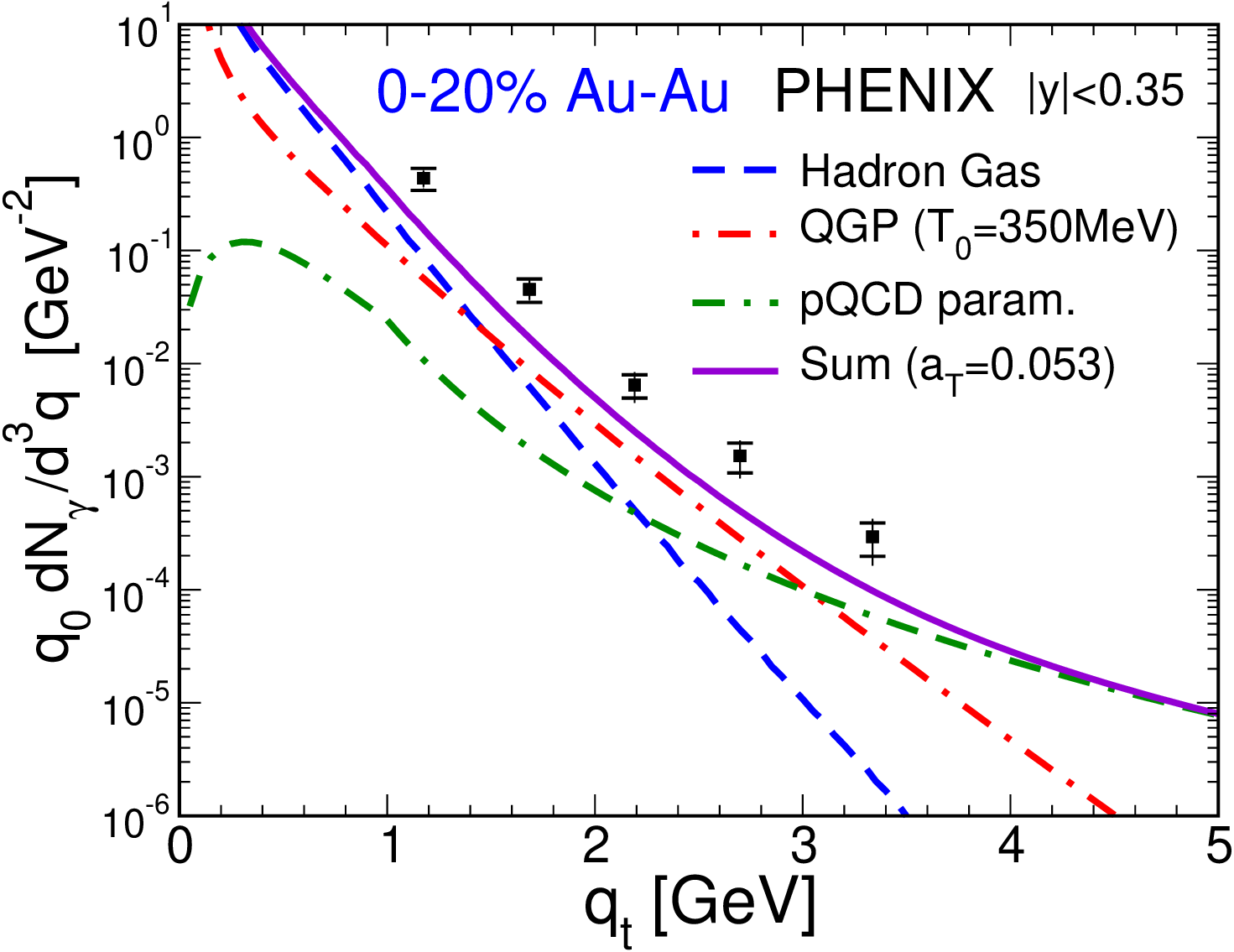}
\end{minipage} \hfill
\begin{minipage}{0.9\linewidth}
\includegraphics[width=\textwidth]{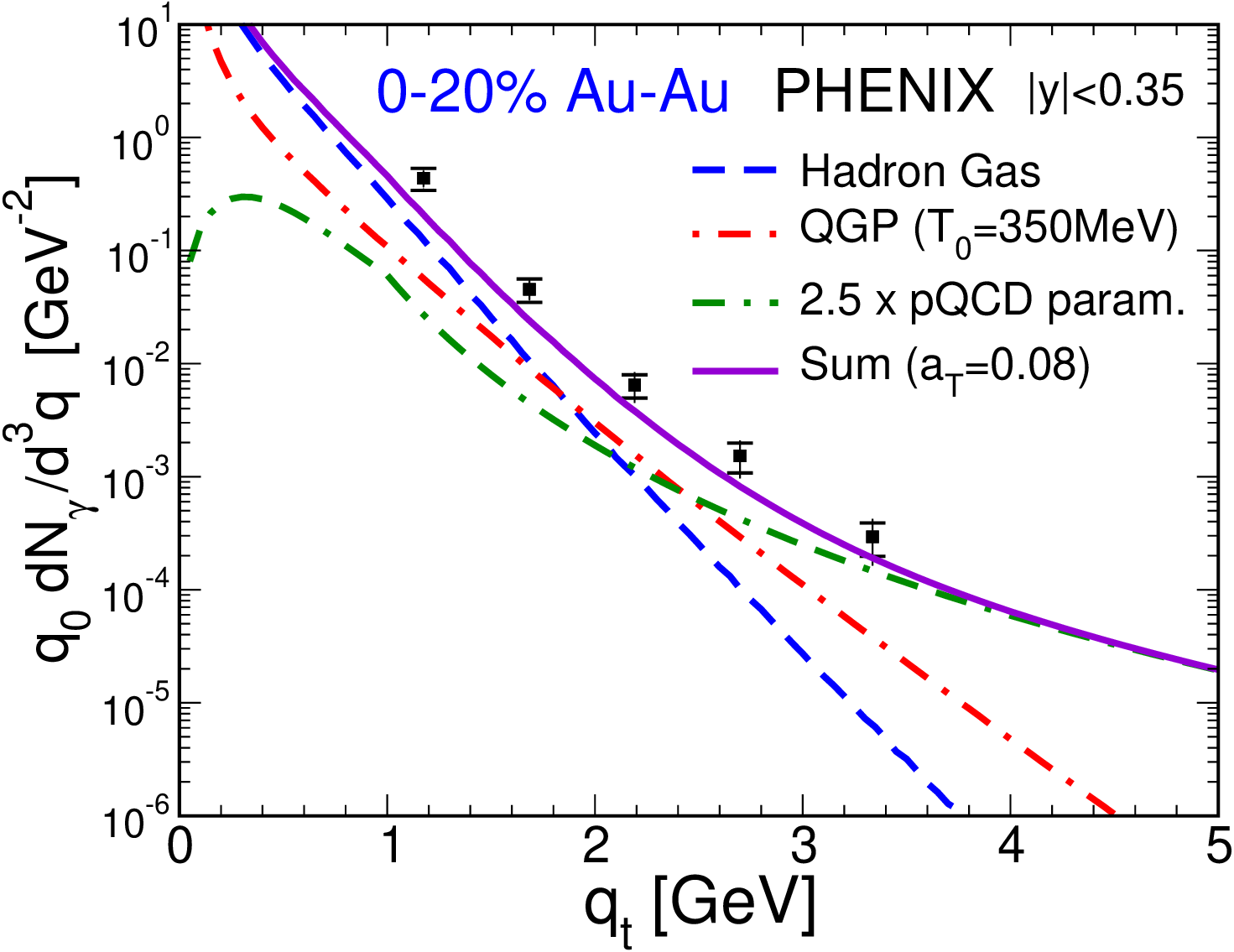}
\end{minipage}\hfill
\begin{minipage}{0.9\linewidth}
\includegraphics[width=\textwidth]{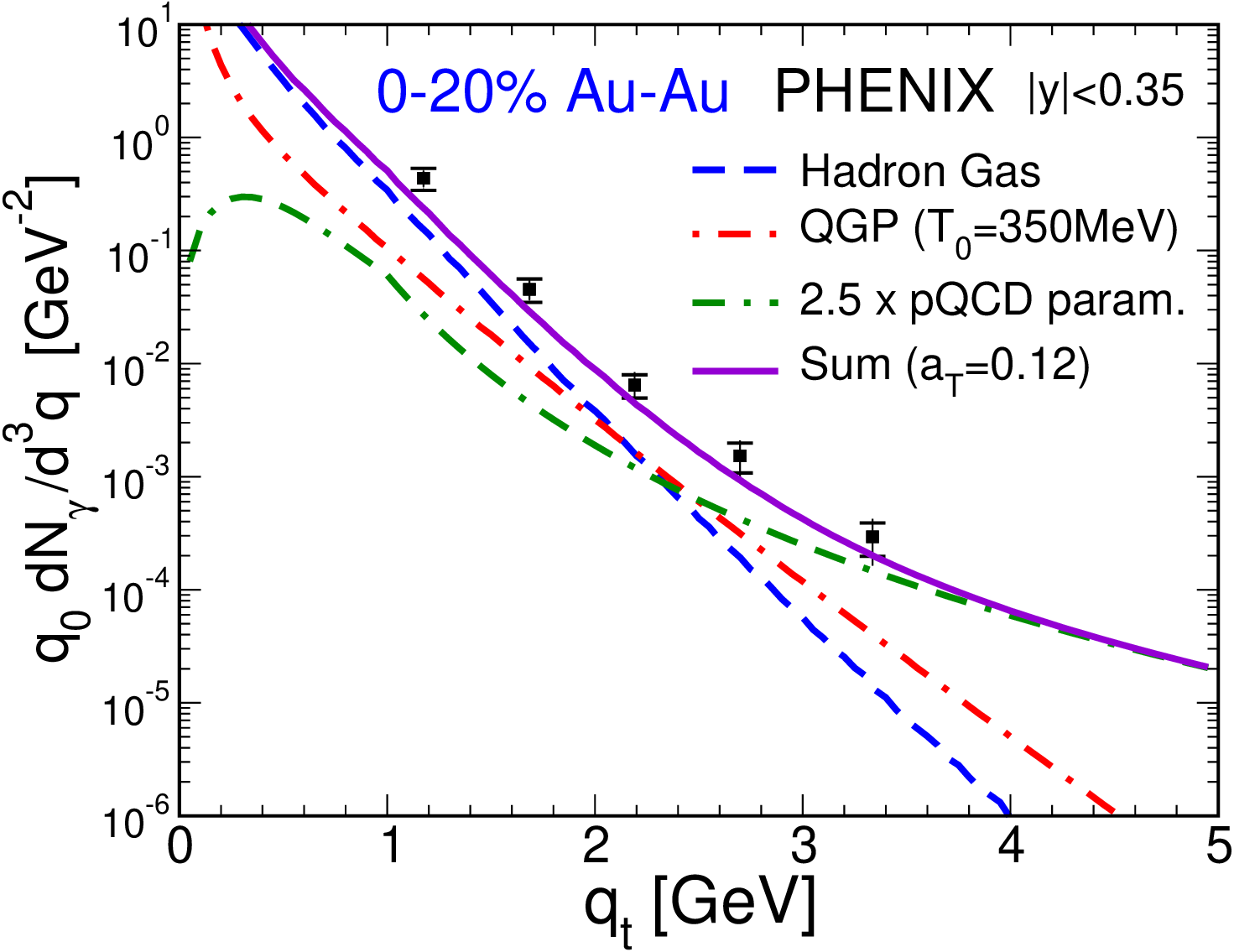}
\end{minipage}
\end{center}
\caption{(Color online) The impact of an increasingly strong radial flow
  in an expanding fireball model on direct-photon spectra in 0-20\%
  central Au-Au collisions at RHIC. The transverse fireball acceleration
  increases from $0.053/\fm$ (upper panel, corresponding to
  Ref.~\cite{Turbide:2003si}) via 0.08 (middle panel; see
  Refs.~\cite{vanHees:2007th,Rapp:2009yu}) to 0.12/fm (lower panel).
  The same QGP and hadronic emission rates have been used in all cases,
  while the primordial contribution has been upscaled in the middle and
  lower panel. As a benchmark, we also show the pertinent PHENIX
  data~\cite{Adare:2008fq}.  }
\label{fig_ph-at}
\end{figure}
The pertinent photon spectra, displayed in the upper panel of
Fig.~\ref{fig_ph-at}, closely resemble the results of Fig.~12 in
Ref.~\cite{Turbide:2003si}.\footnote{We note that in Figs.~12 and 13 of
  Ref.~\cite{Turbide:2003si} the contribution labeled ``Hadron Gas''
  only includes the in-medium $\rho$ spectral function part, not the
  meson-gas contributions also calculated in there. Unfortunately, we
  recently realized that the spectral function part in the photon
  spectra at RHIC and LHC (Figs.~12 and 13 in
  Ref.~\cite{Turbide:2003si}) was computed with the spin-averaged $\rho$
  propagator, $D_\rho = (2D_\rho^T+D_\rho^L)/3$, which, at the photon
  point (where the transverse part, $D_\rho^T$, should be used), is by a
  factor of 2/3 too small. It was done correctly in the rate plots and
  for the SPS calculations shown in Ref.~\cite{Turbide:2003si}. In the
  present work we refer to the ``Hadron Gas'' emission as the sum of
  spectral-function and meson-gas contributions.}
\begin{figure*}[!t]
\begin{center}
\begin{minipage}{0.49\linewidth}
\includegraphics[width=1.03\textwidth]{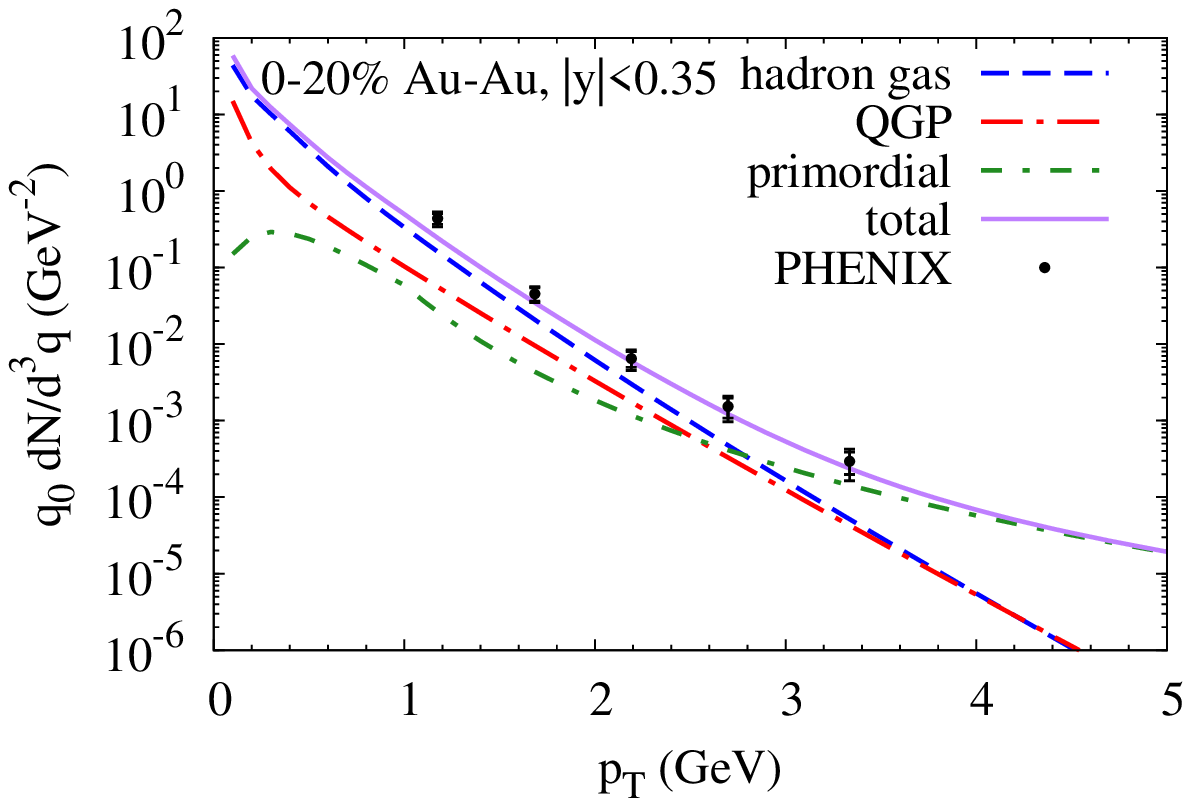}
\end{minipage} \hfill
\begin{minipage}{0.49\linewidth}
\includegraphics[width=1.06\textwidth]{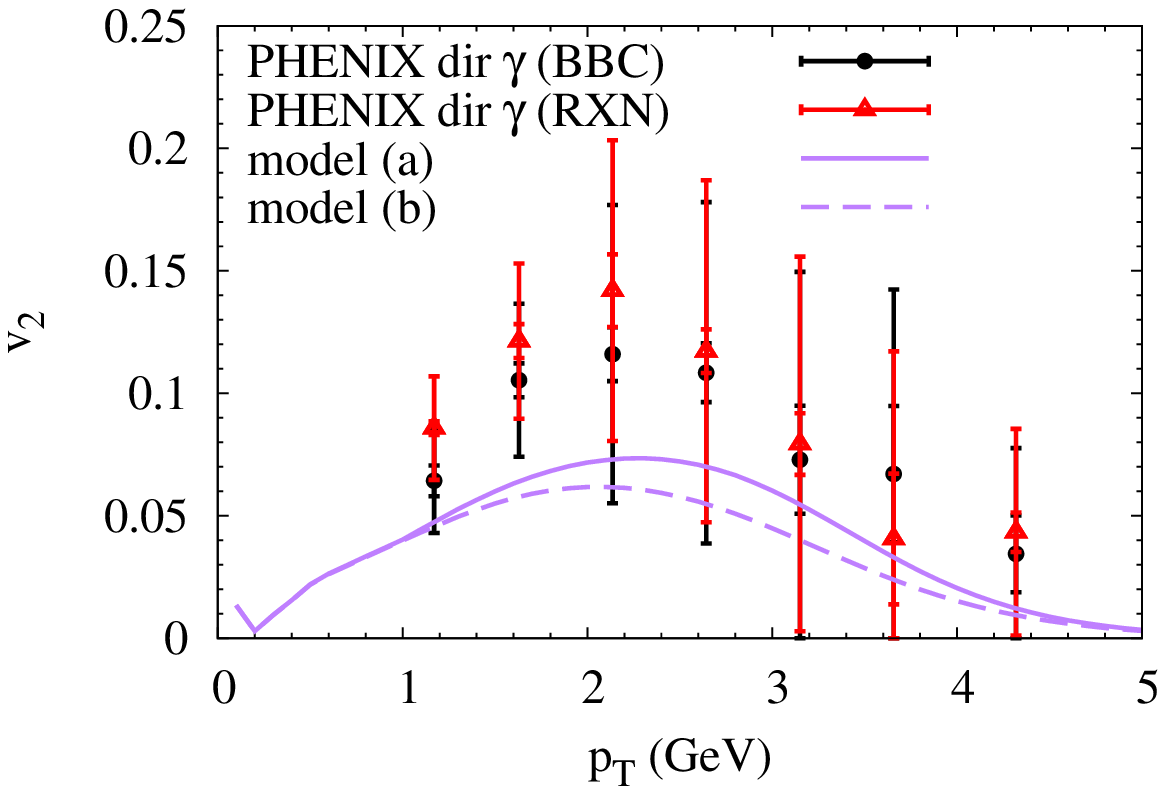}
\end{minipage}
\begin{minipage}{0.48\linewidth}
\includegraphics[width=\textwidth]{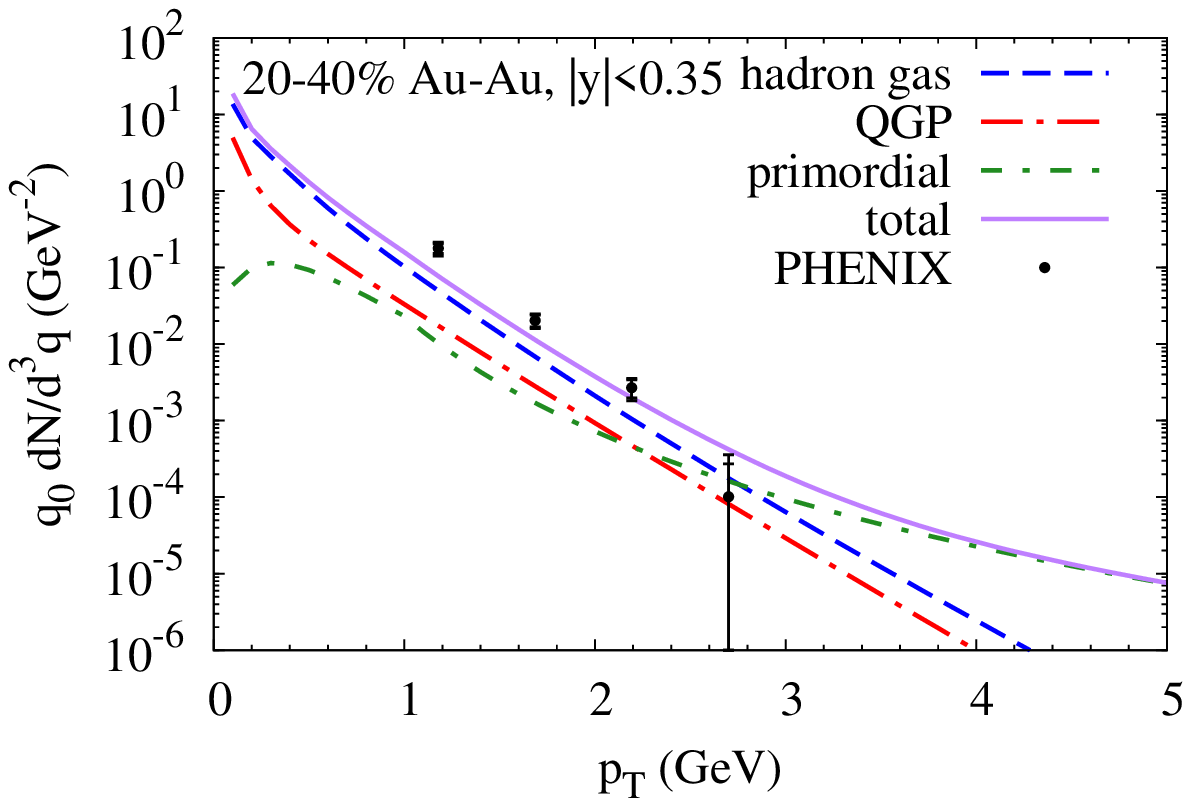}
\end{minipage} \hfill
\begin{minipage}{0.48\linewidth}
\includegraphics[width=\textwidth]{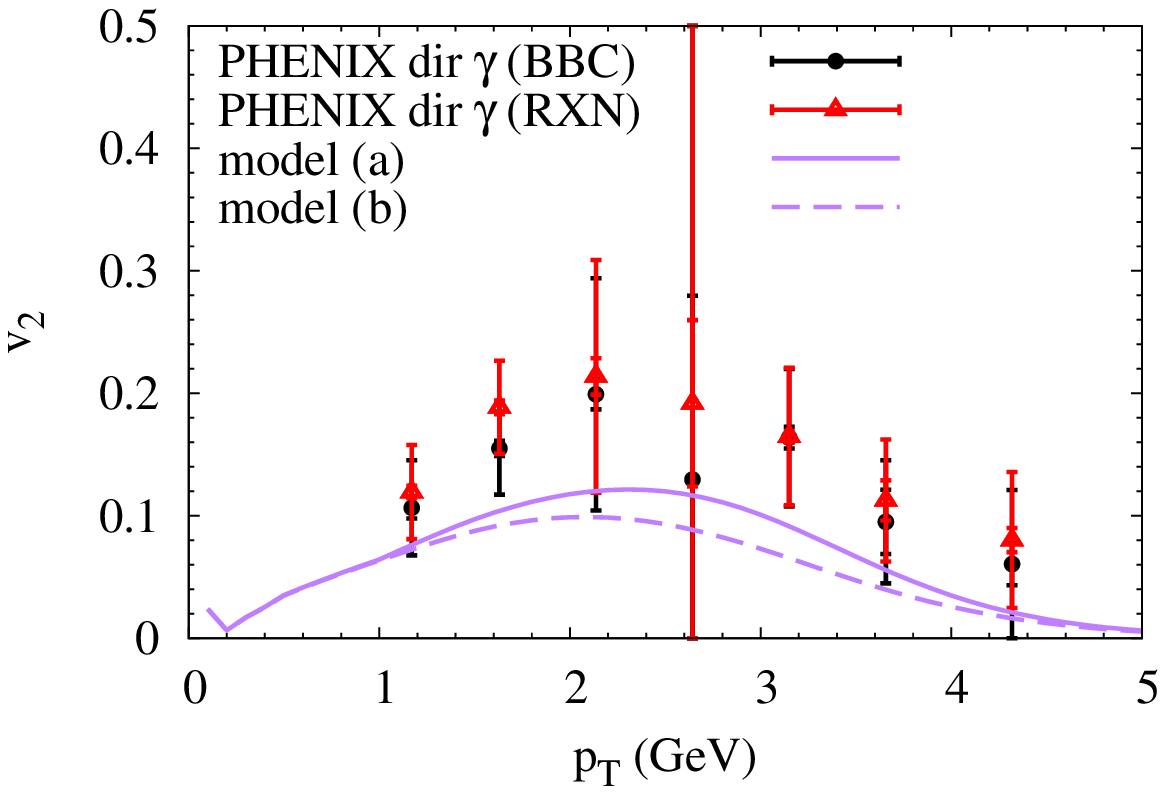}
\end{minipage}
\end{center}
\caption{(Color online) Comparison of our calculated direct-photon
  spectra (left panels) and their elliptic-flow coefficient (right
  panels) from an elliptically expanding fireball model with QGP and
  hadronic radiation, supplemented with primordial emission, to PHENIX
  data~\cite{Adare:2008fq,Adare:2011zr} in 0-20\% (upper panels) and
  20-40\% (lower panels) central Au-Au($\sqrt{s}=200 \; A \GeV$)
  collisions.  Models (a) and (b) in the right panels refer to the use
  of the pQCD parameterization and the PHENIX fit for primordial
  production, respectively (in the left panels, only model (a) is
  displayed).}
\label{fig_ph-qt} 
\end{figure*}
A window of QGP-radiation dominance is present for $q_t\simeq1.5$-$3\;
\GeV$.

The situation changes somewhat with an update performed in 2007
triggered by the analysis of NA60 dileptons at the SPS, specifically in
the context of their $q_t$-spectra. The fireball acceleration was
increased to $a_T=0.08$-$0.1 c^2/\fm$ to better reproduce hadron
spectra, which also allowed for a significantly improved description of
the slope parameters in the dilepton $q_t$
spectra~\cite{vanHees:2007th}. It was also checked that the agreement
with the WA98 direct-photon spectra at SPS~\cite{Aggarwal:2000th}, as
found in Ref.~\cite{Turbide:2003si}, was not distorted (see, \eg,
Fig.~23 in Ref.~\cite{Rapp:2009yu}). At RHIC, the pertinent fireball of
lifetime $\tau\simeq15 \; \fm/c$ results in a freezeout temperature of
$T_{\mathrm{fo}}=98\; \MeV$ with a surface transverse flow of
$\beta_s=0.77$. The consequences for the direct-photon spectra, using
the same thermal emission rates and fireball chemistry as before, are
illustrated in the middle panel of Fig.~\ref{fig_ph-at}: while the
spectral distribution of the QGP radiation is barely affected, the
hadronic radiation spectrum becomes noticeably harder, thus shifting the
crossing with the QGP part up to $q_t\simeq1.8 \; \GeV$. In combination
with an improved estimate of the primordial emission, adjusted to then
available PHENIX $pp$ data, the QGP window shrinks appreciably, with a
maximum fraction of ca.~42\% of the total at $q_t\simeq2.1 \; \GeV$.

Finally, recent systematic analyses of light-hadron spectra by the STAR
collaboration~\cite{Abelev:2008ez} requires an even harder expansion, to
reach a thermal freezeout configuration at $T_{\mathrm{fo}}\simeq
90$-$95\; \MeV$ and $\langle\beta\rangle\simeq0.59$ for central Au-Au
collisions. This can be achieved in the fireball model by a further
increase of the acceleration to $a_T=0.12 \; c^2/\fm$ at a slightly
reduced lifetime of $\tau\simeq14 \; \fm/c$. As expected, for the
direct-photon spectra this implies a further hardening of the hadronic
emission and thus an additional squeezing of the QGP window to a small
region around $q_t\simeq2.4\; \GeV$, cf.~lower panel of
Fig.~\ref{fig_ph-at}. The increasing transverse flow of the three
fireballs also seems to improve the description of the PHENIX
direct-photon spectra, although that was not the objective of this
exercise.

\begin{figure*}[!t]
\begin{center}
\begin{minipage}{0.49\linewidth}
\includegraphics[width=1.0\textwidth]{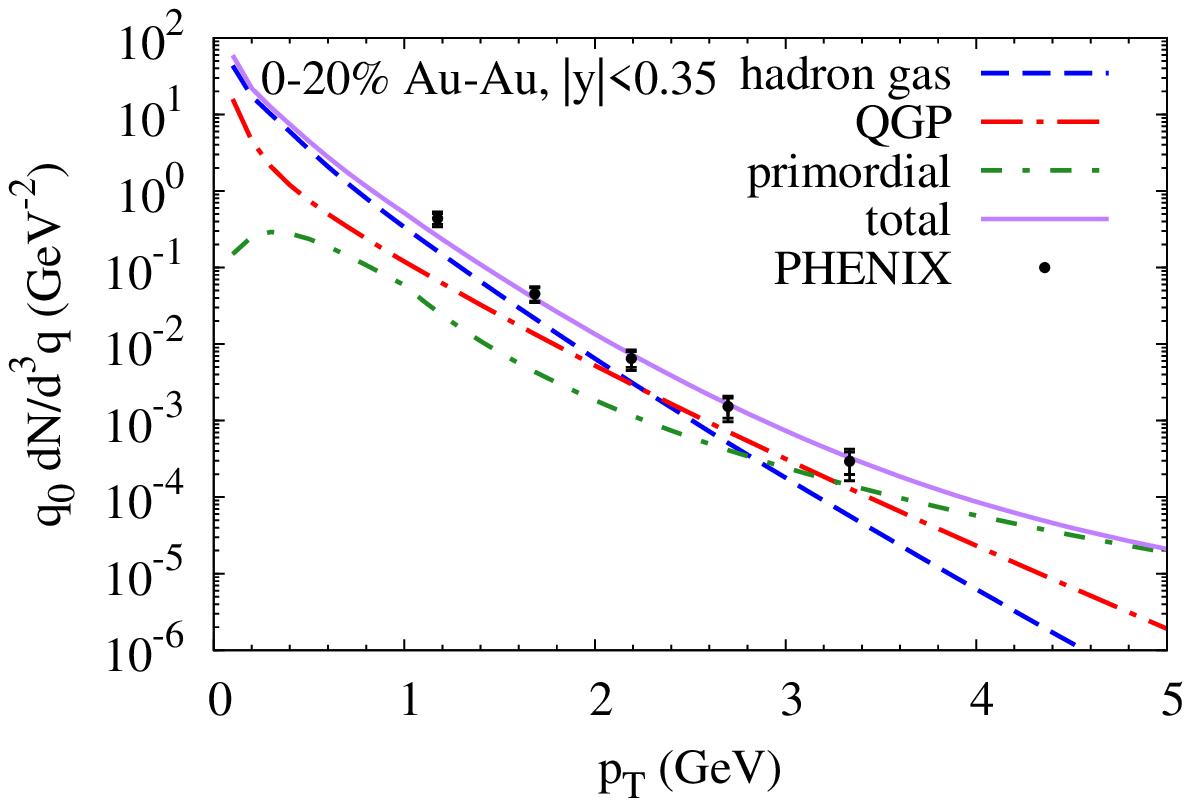}
\end{minipage} \hfill
\begin{minipage}{0.49\linewidth}
\includegraphics[width=1.0\textwidth]{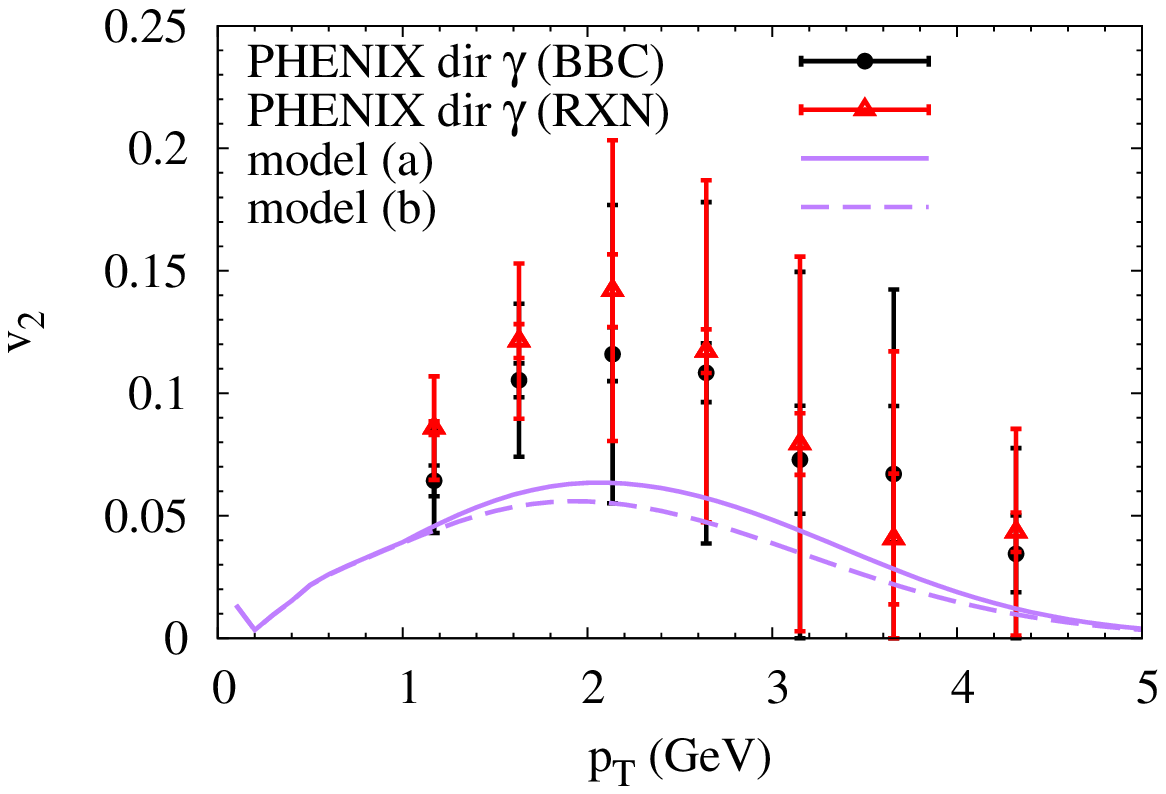}
\end{minipage}
\end{center}
\caption{(Color online) Same as Fig.~\ref{fig_ph-qt} for 0-20\% Au-Au
  collisions, but with a QGP contribution evaluated for a reduced
  thermalization time of $\tau_0\simeq0.17\;\fm/c$ translating into an
  average initial temperature of $T_0\simeq445\; \MeV$.}
\label{fig_ph-tau0}
\end{figure*}

The final (third) fireball setup has been refined by implementing
realistic ellipticities and an explicit linearly increasing flow profile
in the transverse boost of the photon emission rate. The corresponding
comparisons to hadronic data have been discussed in the previous
section.  The resulting direct-photon $q_t$ spectra from thermal QGP and
hadronic sources, supplemented with an $N_{\mathrm{coll}}$-scaled
primordial contribution, are compared to PHENIX data in the 0-20\% and
20-40\% centrality classes of Au-Au($\sqrt{s}=200 A\GeV$) in the two
left panels of Fig.~\ref{fig_ph-qt}. The more central data set is fairly
well reproduced, even though there appears to be a slight
underestimation of the datum at the lowest $q_t\simeq1.2\; \GeV$. The
inclusion of the full transverse-flow profile leads to a further
hardening of the hadronic component relative to the lower panel in
Fig.~\ref{fig_ph-at}, while the QGP component is essentially unaffected,
even at the highest $q_t$ (continuing the constant trend of the three
panels in Fig.~\ref{fig_ph-at}). This means that the high-$q_t$ QGP
radiation is entirely determined by the earliest radiation, where no
flow has built up yet; the subsequent QGP flow cannot overcome the
softening due to the decreasing temperature.  This further implies a
significant dependence of the high-$q_t$ QGP yield on the thermalization
time (a quantitative example will be discussed at the end of
Sec.~\ref{sec_v2}). On the other hand, hadronic emission only sets in at
$T_c$ when there is already substantial flow in the system, and thus
even at high momenta the hadronic spectra are sensitive to the fireball
flow field.

In the 20-40\% centrality bin, the discrepancy between the theoretical
yields and the data becomes somewhat more severe, hinting at a missing
relatively soft source (and therefore suggestive for the later hadronic
phase). One speculation at this point could be related to
$\omega\to\pi^0\gamma$ decays. These have been subtracted by the PHENIX
collaboration employing $m_t$ scaling of the $\omega$ spectra with
$\pi^0$'s~\cite{David:2011priv}, assuming $\omega/\pi^0=1$, as found in
$pp$ measurements~\cite{Adler:2006hy}, as well as in 0-92\% Au-Au
collisions for $p_t>4\; \GeV$. \emph{If}, however, $\omega$ mesons at
lower $p_t$ become part of the chemically equilibrated medium in
heavy-ion collisions, one expects their multiplicity at given $m_t$ to
be up to 3 times larger, due their spin degeneracy. In this case there
might be a direct-photon component in the Au-Au data at low $q_t\le 2 \;
\GeV$ due to some fraction of final-state $\omega\to\pi^0\gamma$ decays
which have not been subtracted (and which would carry large $v_2$). This
possibility may be worth further experimental and theoretical study.

It is quite remarkable that the hadronic yield dominates over the QGP
one over the entire plotted range. This will have obvious ramifications
for the $v_2$ of the direct photons, which is larger in the hadronic
phase. The sub-leading role of the (early) QGP component further implies
that the effects of initial-state fluctuations on thermal-photon
production~\cite{Dion:2011vd,Chatterjee:2011rg} are diminished.

To examine the dependence of the QGP yield on the thermalization time,
we have conducted calculations with a factor-2 reduced initial
longitudinal size, $z_0 =0.3\;\fm$, corresponding to
$\tau_0\simeq0.17\;\fm/c$ as used, \eg, in
Ref.~\cite{Holopainen:2011pd}, cf.~Fig.~\ref{fig_ph-tau0}. The QGP
spectra in 0-20\% Au-Au collisions increase over the $z_0 =0.6\; \fm$
calculation by a factor of 1.6, 2.7 and 4.8 at $q_t=2$, 3 and $4\;\GeV$,
respectively, and turn out to be in fair agreement (within ca.~30\%)
with the hydrodynamic calculations reported in
Ref.~\cite{Holopainen:2011pd} (using smooth initial conditions). The
significance of this increase mostly pertains to momenta, $q_t>2\;
\GeV$, where a small ``QGP window'' reopens, but it does not
significantly affect the description of the experimental yields.

\begin{figure}[!t]
\begin{minipage}{0.95\linewidth}
\includegraphics[width=0.95\textwidth]{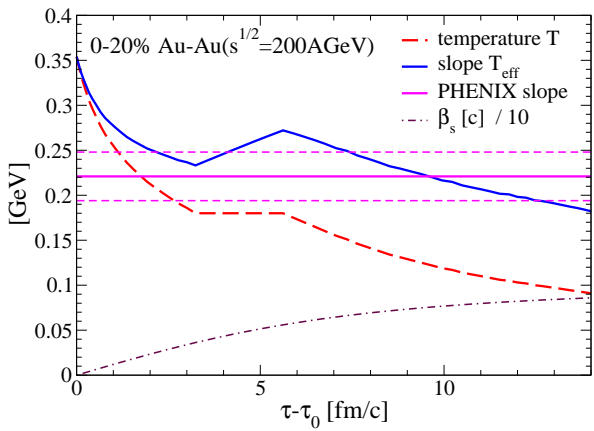}
\end{minipage} \hfill
\caption{(Color online) Time dependence of the effective slope parameter
  of thermal-photon radiation emitted from our fireball for 0-20\% Au-Au
  collisions (solid (blue) line, given by QGP and hadronic sources for
  $\tau-\tau_0\lsim3\; \fm$/$c$ and $\gsim6$\fm/$c$, respectively),
  using Eq.~(\ref{Teff}) with $\langle\beta\rangle=0.7\beta_s$.  For
  comparison, we plot the effective slope extracted by PHENIX from their
  data (horizontal line with dashed lines indicating the experimental
  error)~\cite{Adare:2008fq}, and the ``true" temperature in the thermal
  restframe (long-dashed line).  Also shown is the surface flow velocity
  of the fireball (dash-dotted line).}
\label{fig_slopes}
\end{figure}

To further characterize the nature of the direct-photon excess (i.e.,
beyond the $pp$-scaled primoridial emission), we evaluate the effective
slope parameters, $T_{\mathrm{eff}}$, of our thermal spectra. We recall
that PHENIX extracted the effective slope of the excess radiation in
their data as
$T_{\mathrm{eff}}=221\pm19^{\mathrm{stat}}\pm19^{\mathrm{syst}}\;
\MeV$~\cite{Adare:2008fq}. In Fig.~\ref{fig_slopes} we compare this
range with the temperature evolution, $T(\tau)$, of our fireball; they
only overlap inside the QGP phase. However, when accounting for the
flow-induced blue shift, as estimated by the schematic expression for a
massless particle,
\begin{equation}
T_{\mathrm{eff}} \simeq T
\sqrt{\frac{1+\langle \beta \rangle}{1-\langle \beta \rangle}} \ , 
\label{Teff}
\end{equation}
the overlap with the experimental window is shifted to significantly
later in the evolution, mostly for a flowing hadronic source with a
restframe temperature of $T\simeq 100$-$150\;\MeV$. This suggests a
reinterpretation of the experimental slope as mainly hadronic in origin,
which, as we will see in Sec.~\ref{sec_v2} below, is further supported
by the $v_2$ data. An explicit fit of the slope to our total thermal
spectrum from the elliptic fireball (with $T_0=355\; \MeV$) in the range
$q_t\simeq1-3\; \GeV$ yields $T_{\mathrm{eff}}\simeq240$-$250\; \MeV$,
which is at the upper end of the data (consistent with the slight
underestimate of the lowest-$q_t$ datum; also note that the use of the
average, $\langle \beta \rangle=0.7\beta_s$ in Eq.~(\ref{Teff}), tends
to underestimate the actual slopes, especially at high $q_t$ and
$\beta_s$; we noted that already when going from the spectra in the
lower panel of Fig.~\ref{fig_ph-at} to the full results in the upper
left panel of Fig.~\ref{fig_ph-qt}). Higher initial temperatures are
less favorable, since they result in a further increase of the slope,
\eg, by 10-$15\; \MeV$ for $T_0=445\; \MeV$.

\section{Direct-Photon Elliptic Flow}
\label{sec_v2}

With a fair description of the photon $q_t$ spectra at hand, we proceed
to the calculation of the direct-photon elliptic flow as a function of
transverse momentum, $v_2(q_t)$. The results for the 0-20\% and 20-40\%
centrality classes of Au-Au collisions at RHIC are again compared to
PHENIX data, see upper- and lower-right panel of Fig.~\ref{fig_ph-qt},
respectively. The shape of our calculated $v_2(q_t)$ matches the
measurements rather well, but our maximum value of ca.~7.5\% (12\%) is
below the central values of the 0-20\% (20-40\%) data. Our results are a
factor of 3-4 larger than existing calculations using hydrodynamic
expansions, and reach into the lower end of the (mostly systematic)
experimental uncertainties. The main differences compared to the hydro
calculations are the following: our equilibrium hadronic
rates~\cite{Turbide:2003si} are significantly larger than in previous
studies, our hadronic phase includes meson-chemical potentials and lasts
longer (both due to a smaller $T_{\mathrm{fo}}$ as dictated by data and
a slightly larger $T_{\mathrm{ch}}$ as used in our previous
calculations~\cite{Rapp:2000pe,Rapp:2002fc,Turbide:2003si}), and our
bulk elliptic flow is built up faster than in standard (ideal) hydro
calculations (recall that only a small increase of the $v_2$ during the
hadronic phase facilitates its $\mathrm{KE}_T$-scaling properties of
multi-strange and light hadrons~\cite{He:2010vw}; also note that a more
rapid expansion in the QGP and transition region has been identified as
an important ingredient to solve the ``HBT
puzzle''\cite{Pratt:2008qv}). The combined effect of these four points
is a thermal source which is dominated by hadronic emission carrying
most of the finally observed elliptic flow from its beginning on, i.e.,
for $T\le T_c$.  For the reduced thermalization time of $\tau_0=0.17\;
\fm/c$ the maximal $v_2$ drops by ca.~15\% to $v_2^{\mathrm{max}}\simeq
6.3\%$, cf.~right panel in Fig.~\ref{fig_ph-tau0}.

In the present work, we have assumed a critical temperature of
$T_c\simeq180\; \MeV$, which is in line with $N_f=2+1$ flavor lattice
calculations reported, \eg, in Ref.~\cite{Cheng:2009zi}. However, very
recent lattice data~\cite{Bazavov:2011sd} (as well as earlier
ones~\cite{Borsanyi:2010bp}) indicate that $T_c$ could be as low as
155-$160 \; \MeV$ and therefore one should ask what impact this could
have on our results. A pertinent study has been done in the context of
dilepton production at SPS energies~\cite{vanHees:2007th}. When varying
the critical temperature by $\pm 15\; \MeV$ around the default value of
$T_c=175\; \MeV$, the QGP emission spectra in the intermediate-mass
region ($M\gsim1\; \GeV$) vary by up to $\pm 50\%$ (less at higher
masses where the contribution from earlier phases increases). At the
same time, the hadronic emission part varies by approximately the same
amount in the opposite direction, so that the total yield roughly stays
the same while the relative QGP and hadronic partition varies
appreciably. We expect similar effects for the photon $q_t$ spectra. If
one still requires the multi-strange hadron spectra to freeze out at
$T_c$, we expect that the $v_2$ in the hadronic phase does not
significantly change with $T_c$. However, since for smaller $T_c$ the
hadronic contribution to the direct photon yields is reduced, so should
be the total (weighted) $v_2$. For example, if for the 0-20\% centrality
class the QGP$-$hadronic partition of 1/3$-$2/3 at $q_t\simeq2\; \GeV$
changes to 1/2$-$1/2, we estimate that $v_2(q_t)$ is lowered from 7\% to
6\%.

\section{Conclusions}
\label{sec_concl}

We have updated and extended our calculations of thermal-photon spectra
at RHIC by constructing an improved elliptic fireball expansion which is
quantitatively constrained by bulk-hadron data. In particular, we have
implemented the notion of sequential freezeout by reproducing an
empirical extraction of radial and elliptic flow from multi-strange and
light-hadron data at chemical and kinetic decoupling, respectively. The
fireball evolution has been combined with existing photon emission rates
in the hadronic and QGP phases to obtain thermal-photon spectra in Au-Au
collisions at RHIC. Supplemented with a primordial component estimated
from $pp$ collisions, we have compared our calculations to spectra and
$v_2$ of direct photons as recently measured by the PHENIX
collaboration. Due to a large medium flow (as required by hadron data),
relatively large hadronic photon rates (approximately degenerate with
QGP rates around $T_c$), and effective chemical potentials to conserve
the observed hadron ratios, we have found that the hadronic medium
outshines the QGP for most of the momenta where thermal radiation is
relevant. This, in turn, leads to a ma\-ximal elliptic flow coefficient
of $v_2\simeq 10 \%$ in semicentral Au-Au, which is a factor of $\sim 3$
increased over previous estimates based on QGP-dominated emission.
Consequently, the discrepancy with the PHENIX $v_2$ data is reduced
appreciably. Our results are corroborated by evaluating the effective
slope parameters of the radiation from the thermal source, which have
the largest overlap with the experimental value of
$T_{\mathrm{eff}}\simeq220\; \MeV$ in the (flowing) hadronic
phase. Initial QGP temperatures of above $T_0\simeq400\; \MeV$ are
increasingly disfavored by both the slope and $v_2$ data. Further
scrutiny is needed whether these results can be confirmed in dynamical
space-time models, i.e., in hydrodynamical and transport simulations,
and what the quantitative impact of the transition temperature is
(using, \eg, the (3+1)D viscous hydrodynamics of
Ref.~\cite{Schenke:2010nt}). The prevalence of the hadronic emission in
our calculations reiterates the necessity of a good understanding of the
strongly coupled hadronic phase in heavy-ion collisions. With these
considerations, a satisfactory explanation of the (surprisingly?)
strong direct-photon $v_2$ signal at RHIC might be possible. The
extension of the $v_2$ studies to virtual photons (aka dileptons) would
also be illuminating. First phenomenological studies of this observable
have been initiated~\cite{Chatterjee:2007kx,Deng:2010pq} and are also
planned within our framework.

\acknowledgments 

We thank G.~David and M.~He for valuable discussions. The work of RR is
supported by the US National Science Foundation under grant no.
PHY-0969394 and by the A.-v.-Humboldt foundation. The work of HvH is
supported by the Hessian LOEWE initiative through the Helmholtz
International Center for FAIR. CG is funded by the Natural Sciences and
Engineering Research Council of Canada.

\begin{flushleft}

\end{flushleft}
\end{document}